\documentclass[reprint,aps,prb,amsmath,superscriptaddress,notitlepage,nofootinbib,longbibliography,floatfix]{revtex4-2}
\usepackage[T1]{fontenc}
\usepackage[latin9]{inputenc}

\usepackage{microtype}
\usepackage{newtxtext,newtxmath,xspace}
\usepackage{amsbsy,bm,bbold}
\usepackage{siunitx}
\usepackage{graphicx,color,xcolor,epsfig,rotate}
\usepackage{gensymb}
\usepackage[colorlinks=true,
            linkcolor=blue,
            urlcolor=blue,
            citecolor=blue]{hyperref}
\usepackage{soul}
\usepackage{orcidlink}
\usepackage{longtable}
\usepackage{array}
\usepackage{varwidth}
\usepackage{float}
\usepackage{enumitem}
\newcommand{\iu}{{i\mkern1mu}}

\begin{document}

\title{Symmetry-Adapted Eigensolver from Real-Space Sampling of Tensor-Network-Bridged Quantum Circuits 
}

\author{Lei Xu}
\affiliation{School of Physics, Zhejiang University, Hangzhou 310058, China}

\author{Ling Wang}
\email{mxwang@gmail.com}
\affiliation{School of Physics, Zhejiang University, Hangzhou 310058, China}

\date{\today}

\begin{abstract}
We propose a symmetry-adapted eigensolver for strongly correlated
systems that separates where the wave function is \emph{sampled} from
where it is \emph{defined}. The lattice is divided into equal blocks:
parallel quantum circuits absorb the entanglement within each block,
while a matrix product state (MPS) backbone bridges the blocks and
restores global correlations. The resulting real-space amplitude
$\phi(a_{\rm real})$ acts only as a sample generator---configurations
are drawn from $|\phi(a_{\rm real})|^2$ directly, block by block,
without any Markov chain. The variational state itself is defined in a
symmetry-adapted basis by two separate rules: the modulus of
$\psi(a_{\rm symm})$ is the square root of the total weight
accumulated over the entire equivalence class, $\widetilde{p}(a_{\rm
  symm})=\sum_{a_{\rm real}\in\{g a_{\rm repr}\}}|\phi(a_{\rm
  real})|^2$, whereas its phase is inherited from the single
representative amplitude $\phi(a_{\rm repr})$. The modulus rule is
what makes the scheme both exact and cheap: the distribution
$\widetilde{p}(a_{\rm symm})$ sampled in the symmetric basis is
exactly the one induced by the mapping from the real-space
distribution $p(a_{\rm real})$, so every generated configuration is an
unbiased sample of $|\psi(a_{\rm symm})|^2$; the phase rule then fixes
the only freedom that remains. Because amplitudes within a class are
combined in probability rather than summed with symmetry characters,
$\psi$ is \emph{not} a group projection of the circuit-MPS state; the
class-wide aggregation instead supplies long-range, symmetry-enforced
entanglement that the MPS never has to carry itself. For the periodic
$J_1\!-\!J_2$ antiferromagnetic Heisenberg model with translation,
reflection, inversion, and particle-number symmetries, MPS bond
dimensions up to 6 yield absolute errors in the energy per site of $10^{-5}$ for a
64-site periodic chain and a $6\times6$ torus after bond-dimension
extrapolation.

\end{abstract}

\maketitle

\section{Introduction}

Simulating strongly correlated quantum many-body systems on classical computers remains a formidable challenge due to the exponential growth of the Hilbert space with system size. This vast dimensionality renders brute-force methods such as exact diagonalization (ED) infeasible for large systems~\cite{SandvikAW2010_lecture,VieijraT2020,WeisseA2008a,LaflorencieN2004a,LauchliAM2011}. Tensor network approaches, including matrix product states (MPS), offer an efficient alternative by compressing the wavefunction~\cite{PhysRevLett.93.040502, PhysRevLett.99.220405, VerstraeteCirac2004PEPS, PhysRevLett.93.227205, VerstraeteMurgCirac2008AdvPhys}; however, accurately capturing entanglement beyond one-dimensional systems requires rapidly increasing bond dimensions. The resulting memory and computational costs quickly become prohibitive, underscoring the need for novel strategies to efficiently represent quantum correlations.

Quantum computation offers a promising route to alleviate the classical bottlenecks in simulating quantum systems~\cite{BiamonteJ2017,PreskillJ2018,CerezoM2021}. A particularly appealing approach is to partition large quantum circuits into smaller subcircuits that can be executed in parallel across multiple quantum processing units. In such a distributed architecture, quantum and classical resources operate synergistically: quantum subcircuits capture local correlations, while classical tensor-network coordination integrates their outputs to reconstruct the global quantum state~\cite{BravyiS2016,EddinsA2022,NiuJ2023,CarreraVazquezA2024,EndoS2021}. Although this collective quantum-classical workflow effectively addresses the scaling challenge associated with qubit numbers, its ability to faithfully represent strongly correlated systems with substantial entanglement remains limited. Nevertheless, it provides a viable and efficient framework for moderately correlated or weakly entangled problems.

Recently, there has been growing interest in combining tensor-network structures with quantum-circuit representations to leverage the complementary advantages of classical tensor contractions and quantum-state preparation. Hybrid tensor networks use quantum subcircuits to represent local quantum states while connecting them through classically contractible tensor networks, enabling the simulation of larger systems with fewer quantum resources~\cite{Yuan2021}. Quantum-circuit tensor networks replace the dense tensors in conventional MPS or MERA ansatzes with structured local circuits, providing compact and expressive variational states for many-body ground states~\cite{Haghshenas2022}. Similar concepts have been applied to represent Schmidt decompositions, where unitary tensor networks generate Schmidt states and an MPS encodes the Schmidt coefficients, offering an efficient representation of entanglement spectra for bipartitions with nontrivial boundaries~\cite{Zhou2023}.

Our work builds on this direction, but differs from it in the role that the tensor-network-bridged circuit plays. In Refs.~\cite{Yuan2021,Haghshenas2022,Zhou2023} the combined circuit--tensor-network object \emph{is} the variational state, and observables are evaluated on it directly. Here it serves only as a sample generator. Within each block a quantum circuit $C_i$ captures the local entanglement, while the MPS backbone $B_i$ bridges the blocks and restores the correlations between them; the two are not separate ingredients but a single composite ansatz---the tensor-network-bridged quantum circuits---whose amplitude we denote $\phi(a_{\rm real})$. It is this composite object that is sampled: real-space configurations are drawn directly from $|\phi(a_{\rm real})|^2$, block by block~\cite{FerrisAJ2012,HanZY2018}, while the variational state itself is defined in a symmetry-adapted basis built from the full set of lattice symmetries. Symmetry-adapted bases have of course been used before to constrain variational wave functions and to target specific momentum and parity sectors~\cite{ChooK2018,ChooK2019,BaoST2025,Bukov_21}; in those constructions the symmetric state is obtained by projection, so that the amplitudes of symmetry-related configurations are summed coherently and interfere. Our construction is not a projection. Instead, each sampled configuration $a_{\rm real}$ is mapped deterministically onto the representative $a_{\rm repr}$ of its equivalence class, and thereby onto a single symmetric basis state $a_{\rm symm}$; because this map is many-to-one and the configurations are drawn from $p(a_{\rm real})$, the symmetric states so obtained occur with precisely the weight aggregated over the whole class. Within one equivalence class the configurations are therefore combined in probability rather than in amplitude, and the phase of the symmetric component is inherited from its representative configuration alone. This choice is what makes the two halves of the scheme fit together: it renders the distribution sampled in the symmetric basis identical to the one induced by the real-space distribution, so that configurations generated in real space are already unbiased samples of the symmetric wave function, while the class-wide aggregation supplies the long-range entanglement that the MPS backbone would otherwise have to carry.

Concretely, the entire construction follows from a single condition on the
many-to-one mapping between the real-space and symmetric bases,
\begin{equation}
    \widetilde{p}(a_{\text{symm}}) = \sum_{a_{\rm real} \in \{g a_{\rm repr}\}} p(a_{\rm real}),
    \label{pmap}
\end{equation}
where $p(a_{\text{real}})$ and $\widetilde{p}(a_{\text{symm}})$ denote the distributions in the
real-space and symmetric bases, respectively, and $\{g a_{\rm repr}\}$ is the equivalence class of
real-space configurations obtained by applying all symmetry-group elements $g \in \mathcal{G}$ to the
representative configuration $a_{\text{repr}}$, in one-to-one correspondence with the symmetric
state $a_{\text{symm}}$. Read as a definition of $|\psi(a_{\rm symm})|^2$, Eq.~\eqref{pmap} fixes
the modulus of the symmetric amplitude and leaves the phase as the only remaining freedom; we
assign that phase from the representative configuration alone, as specified in
Eqs.~\eqref{amp_psi} and~\eqref{arg_psi}. Two consequences follow. First,
$\widetilde{p}(a_{\rm symm})$ needs no separate normalization, since summing Eq.~\eqref{pmap} over
all symmetric states merely resums $p(a_{\rm real})$ over the complete real-space basis: it is
exactly the distribution induced by the deterministic map $a_{\rm real}\mapsto a_{\rm symm}$, so a
configuration drawn from $p(a_{\rm real})$ is already an unbiased sample of
$\widetilde{p}(a_{\rm symm})$, requiring neither reweighting nor Markov-chain updates. Second,
because every symmetric basis state superposes configurations spread across the entire lattice,
the class-wide aggregation endows $\psi$ with long-range entanglement that the real-space MPS
never has to carry, which is what permits the very small bond dimensions used below.

To validate the effectiveness of the proposed eigensolver, we apply it to the periodic $J_1\!-\!J_2$ antiferromagnetic Heisenberg model in one and two dimensions--a benchmark system for studying frustration and quantum correlations. In our demonstration, translation, reflection, inversion, and particle number conservation symmetries are incorporated to ensure that both sampling and measurement procedures fully respect the system's inherent symmetries. With a small matrix product state bond dimension of up to 6, the method achieves an absolute error in the energy per site of $10^{-5}$ for a 64-site periodic chain and a $6\times6$ torus after bond-dimension extrapolation. These results confirm the accuracy and consistency of the approach, while indicating strong potential for scaling to larger and more complex quantum systems in future studies.

The remainder of this paper is organized as follows. In Sec.~\ref{sec:method}, we detail the methodology of the symmetry-adapted eigensolver, including the wave-function construction, the real-space sampling procedure, the energy gradient evaluations, and the quantum circuits. In Sec.~\ref{sec:results}, we present the numerical results for the one- and two-dimensional $J_1\!-\!J_2$ Heisenberg models. Finally, a summary is given in Sec.~\ref{sec:conclusions}. Additional technical details regarding the energy-gradient derivation, the construction of symmetric states, the truncation of the RDM subspace, and the entanglement-entropy comparison are provided in Appendices~\ref{app:energy}, \ref{app:sps}, \ref{app:weight}, and~\ref{app:ee}, respectively.

\section{Method}
\label{sec:method}
\subsection{Wavefunction}
We partition an $N$-site system into $n_b$ blocks, each containing $b$ spins, such that $N = b n_b$. A real-space configuration is denoted as $a_{\rm real} = \otimes_{i=1}^{n_b} \bar{a}_i$, where each block state $\bar{a}_i=\otimes_{k=(i-1)b}^{ib-1}s_{k}$. The real-space amplitude $\phi(a_{\rm real})$, whose modulus squared defines the distribution $p(a_{\rm real}) \equiv |\phi(a_{\rm real})|^2$, is produced by a single composite ansatz whose two parts play complementary roles: a set of parallel quantum circuits $\{C_i\}$ acting within the blocks, and an MPS backbone $\{B_i\}$ bridging them. Each circuit $C_i$ compresses the local spin block $\bar{a}_i$ onto a renormalized basis $\gamma_i$, while the MPS tensors $\{B_i\}$ connect these renormalized bases $\{\gamma_1, \ldots, \gamma_{n_b}\}$ across blocks to produce the amplitude
\begin{align}
    \label{phia}
    &\phi(a_{\rm real})= \nonumber\\
    &\sum_{\{\mu_i,\gamma_i\}} B_{\mu_{1}}^{\gamma_1}C_{\gamma_1}[\bar{a}_1]B_{\mu_{1}\mu_{2}}^{\gamma_2}C_{\gamma_2}[\bar{a}_2]\cdots B_{\mu_{n_b-1}}^{\gamma_{n_b}}C_{\gamma_{n_b}}[\bar{a}_{n_b}].
\end{align}

\begin{figure}
    \centering
    \includegraphics[width=\columnwidth]{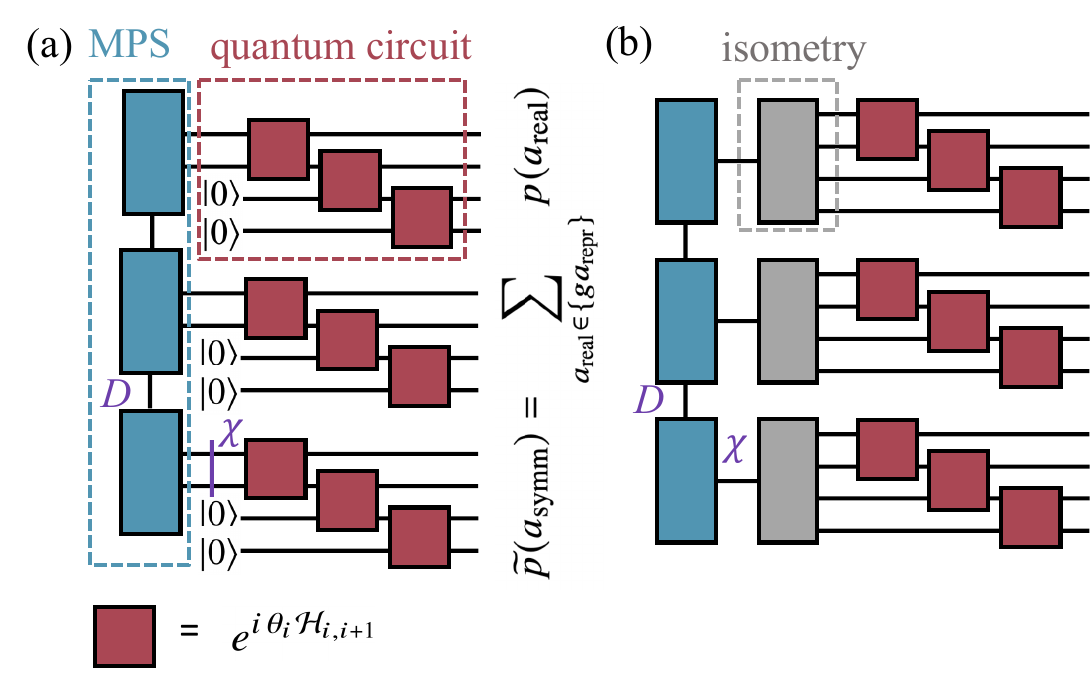}
    \caption{Wavefunctions of the Symmetry-Adapted Eigensolver. (a) Tensor-network-bridged parallel quantum circuits efficiently generate real-space samples $a_{\rm real}$ according to the distribution $p(a_{\rm real})$, ensuring that the corresponding symmetric components $a_{\rm symm}$ follow $\widetilde{p}(a_{\rm symm}) = \sum_{a_{\rm real} \in \{g a_{\rm repr}\}} p(a_{\rm real})$. Here, $D$ denotes the MPS bond dimension, and $\chi$ represents the renormalized block degrees of freedom. (b) An effective isometric transformation can be inserted in place of the quantum circuits, constructed from the dominant eigenvectors of the reduced density matrix obtained from an exactly solvable system.}
    \label{generator}
\end{figure}

This composite structure---the tensor-network-bridged quantum circuits---is illustrated in Fig.~\ref{generator}(a), where each $C_i$ is a local isometry satisfying $C_i^{\dagger} C_i = I_{\chi\times \chi}$, with $\chi \le 2^b$ the number of retained renormalized states. We emphasize that $\phi(a_{\rm real})$ is not itself the variational wave function: it serves solely as a generator of real-space configurations, while the variational state is defined separately in the symmetric basis introduced below.

To define the target symmetric subspace and corresponding wave function, we first construct an orthogonal basis from symmetry-constrained superpositions of the Ising configurations $|a_{\rm real}\rangle = |s_1 \otimes s_2 \otimes \cdots \otimes s_N\rangle$. Taking one-dimensional translational symmetry as an example, a symmetric basis vector is defined as~\cite{BaoST2025,SandvikAW2010_lecture,VieijraT2020,WeisseA2008a,LaflorencieN2004a,LauchliAM2011}
\begin{equation}
|a_{\rm symm}\rangle = \frac{1}{\sqrt{N_{a_{\rm repr}}}}
\sum_{r = 0}^{N - 1} e^{-ikr}  \mathcal{T}^r |a_{\rm repr}\rangle,
\label{mbasis}
\end{equation}
where $\mathcal{T}$ denotes the lattice translation operator and $N_{a_{\rm repr}}$ is a normalization factor, such that $\langle a_{\rm symm}|a_{\rm symm}\rangle=1$. Each representative configuration $|a_{\rm repr}\rangle$ is the lexicographically smallest element within the equivalence class $\{\mathcal{T}^r |a_{\rm repr}\rangle\}$, where spins are encoded in binary form (e.g., $0$ for $\vert\uparrow\rangle$, $1$ for $\vert\downarrow\rangle$).

For a given momentum sector $k$, a state $|\psi\rangle$ satisfying $\mathcal{T}|\psi\rangle = e^{ik}|\psi\rangle$ can be expanded as
\begin{equation}
    |\psi\rangle = \sum_{a_{\rm symm}} \psi(a_{\rm symm}) |a_{\rm symm}\rangle,
\label{wavefunc1}
\end{equation}
where $N_\text{repr}$ is the dimension of the symmetric subspace, which is substantially smaller than the full Hilbert space due to the imposed symmetry and orthogonality constraints. This construction can be straightforwardly generalized to higher dimensions and other symmetry groups, which is illustrated in Appendix~\ref{app:sps}.

It remains to specify the coefficients $\psi(a_{\rm symm})$ of Eq.~\eqref{wavefunc1}. Rather than obtaining them by projecting the generator $\phi(a_{\rm real})$ of Eq.~\eqref{phia} onto the symmetric sector, we define them directly, through two separate rules for the modulus and the phase,
\begin{subequations}
    \begin{align}
	    \Re \ln \psi(a_{\rm symm}) &\equiv \frac{1}{2}\ln \sum_{a_{\rm real}\in\{g a_{\rm repr}\}}|\phi(a_{\rm real})|^2, \label{amp_psi} \\
	    \Im \ln \psi(a_{\rm symm}) &\equiv \Im \ln \phi(a_{\rm repr}),
        \label{arg_psi}
    \end{align}
\end{subequations}
where the summation in Eq.~\eqref{amp_psi} runs over all distinct real-space configurations $a_{\rm real}$ within the equivalence class $\{g a_{\rm repr}\}$, with $g$ enumerating the symmetry-group elements. In this construction, the amplitude of $|a_{\rm symm}\rangle$ equals the square root of the total probability weight aggregated from all configurations in the class, while its phase is inherited from the representative configuration. Individual real-space configurations thus have no independent meaning once mapped into the symmetric basis.

The symmetric wave function is thus tied directly to the generator: the distribution $p(a_{\rm real})$ produced by the tensor-network-bridged circuits fixes the weight of every symmetric basis component. Because the aggregation in Eq.~\eqref{amp_psi} acts on probabilities rather than on amplitudes, phases within a class never interfere and are simply absent from $|\psi(a_{\rm symm})|$, while the relative phases between distinct representatives $a_{\rm repr}$ are retained through Eq.~\eqref{arg_psi}. Owing to this close correspondence between the generator and the symmetric wave function, it is advantageous to initialize the MPS backbone from one optimized within a conventional real-space algorithm, which provides a good starting point for the subsequent optimization.

\subsection{Sampling}
\begin{figure*}
\centering
    \includegraphics[width=\textwidth]{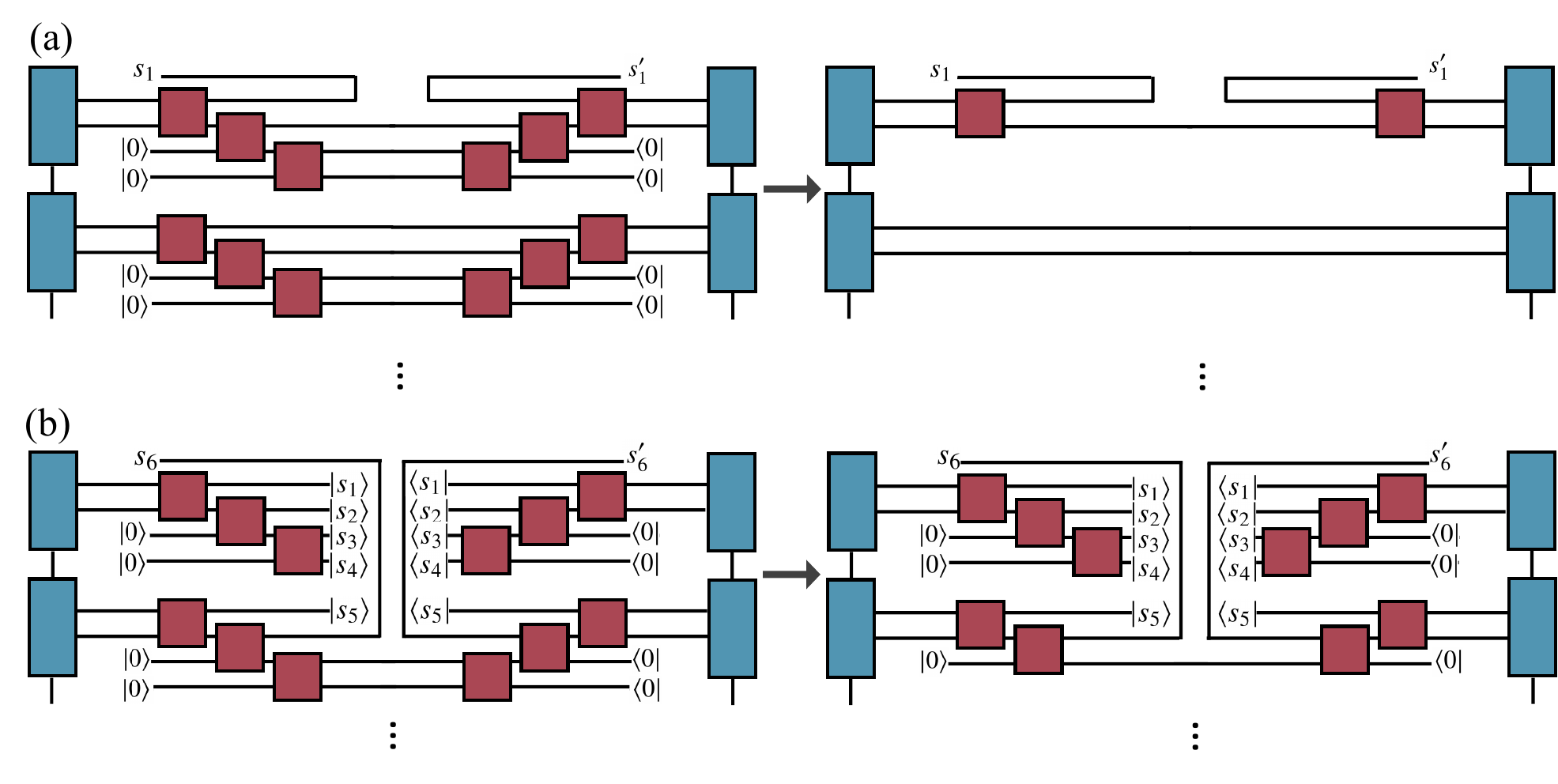}
    \caption{The direct (perfect) sampling algorithm for real-space configurations $a_{\rm real}$. (a) Contraction of the RDM $\rho(s_{1}, s_{1}')$, whose diagonal elements determine the sampling probabilities $p(\uparrow_{1}) = \rho(\uparrow_1,\uparrow_1)/[\rho(\uparrow_1,\uparrow_1) + \rho(\downarrow_1,\downarrow_1)]$ and $p(\downarrow_1) = 1 - p(\uparrow_1)$. (b) Contraction of the conditional RDM $\rho(s_{i}, s_{i}' \,|\, s_{i-1}, \cdots, s_{1})$, conditioned on the previously sampled spins. In both panels, the right-hand diagrams illustrate the simplification arising from the isometric contraction of the quantum circuits, $C_{i}^{\dagger} C_{i} = I_{\chi \times \chi}$.}
    \label{sampling}
\end{figure*}

One major advantage of the construction in Eqs.~\eqref{amp_psi} and~\eqref{arg_psi} is that $\psi(a_{\rm symm})$ can be sampled directly, with no Markov chain, by perfect (direct) sampling of the generator $\phi(a_{\rm real})$~\cite{FerrisAJ2012,HanZY2018}; the same conditional reduced-density-matrix machinery also underlies the minimally entangled typical thermal states (METTS) algorithm at finite temperature~\cite{WhiteSR2009,StoudenmireEM2010,ChungCM2019}. This efficiency arises because a properly generated real-space configuration $a_{\rm real}$, drawn according to the distribution $p(a_{\rm real})=|\phi(a_{\rm real})|^2$, can be directly mapped to its representative configuration $a_{\rm repr}$ and hence to the corresponding symmetric basis component $a_{\rm symm}$ with probability $\widetilde{p}(a_{\rm symm})=\sum_{a_{\rm real} \in \{g a_{\rm repr}\}} p(a_{\rm real})$. Fig.~\ref{sampling} illustrates how $a_{\rm real}$ is generated block by block by direct sampling. As in Fig.~\ref{sampling}(a), the reduced density matrix (RDM) $\rho(s_1,s_1^{\prime})$ can be computed efficiently by tracing out all other degrees of freedom, owing to the isometric property $C_i^{\dagger} C_i = I_{\chi\times\chi}$ of the quantum circuits and the efficient contraction properties of the MPS backbone. From RDM $\rho(s_1,s_1^{\prime})$, the local sampling probabilities are obtained as
\begin{equation}
    p(\uparrow_1)=\frac{\rho(\uparrow_1,\uparrow_1)}{\rho(\uparrow_1,\uparrow_1)+\rho(\downarrow_1,\downarrow_1)}
\end{equation}
where $p(\downarrow_1)=1-p(\uparrow_1)$. Fig.~\ref{sampling}(b) shows how the conditional RDM for the $i^{\rm th}$ spin, $\rho(s_i,s_i^{\prime}|s_{i-1},\ldots,s_1)$, is updated once the preceding spins have been sampled. After a full configuration $a_{\rm real}$ is generated, it is mapped to its corresponding symmetric component $a_{\rm symm}$ for energy and gradient evaluations. Since each real-space configuration is sampled according to its correct statistical weight, and the map from $a_{\rm real}$ to $a_{\rm symm}$ is deterministic and many-to-one, the symmetric components so obtained follow exactly the target distribution $\widetilde{p}(a_{\rm symm})$.

\subsection{Energy and Energy Gradient}
The energy expectation value is evaluated by direct sampling of the symmetric components $a_{\rm symm}$ from $\widetilde{p}(a_{\rm symm}) = |\psi(a_{\rm symm})|^2$; as established above, this distribution is generated exactly by drawing $a_{\rm real}$ from $p(a_{\rm real})$ and mapping each configuration onto its representative. The expectation value is then
\begin{align}
  \langle E\rangle &= \frac{\langle \psi | \mathcal{H} | \psi \rangle} {\langle \psi | \psi \rangle} \nonumber \\
  &=\frac {\sum_{a_{\rm symm}}|\psi(a_{\rm symm})|^2E_{\rm loc}(a_{\rm symm})} {\sum_{a_{\rm symm}}|\psi(a_{\rm symm})|^2},
\end{align}
where
\begin{equation}
    E_{\rm loc}(a_{\rm symm}) = \sum_{a^{\prime}_{\rm symm}}\frac{\psi(a^{\prime}_{\rm symm})}{\psi(a_{\rm symm})}\langle a_{\rm symm}|\mathcal{H}|a^{\prime}_{\rm symm}\rangle.
\end{equation}

The energy gradient, defined as $\nabla_{\theta} E=\partial E / \partial \theta^*$, can be written as
\begin{equation}
\frac{\partial E}{\partial \theta^*} = \frac{\partial}{\partial \theta^*} \frac{\langle \psi | H | \psi \rangle}{\langle \psi | \psi \rangle}.
\end{equation}

By substituting Eqs.~\eqref{amp_psi} and \eqref{arg_psi}, we obtain
\begin{align}
  	\label{egradient}
	\frac{\partial E}{\partial \theta^*} &= \Big\langle \Big( \frac{\partial \Re\ln \phi(a_{\rm real})}{\partial \theta^*} - i\frac{\partial \Im \ln \phi(a_{\rm repr})}{\partial \theta^*} \Big) E_{\rm loc}(a_{\rm symm}) \Big\rangle_{a_{\rm real} } \nonumber \\
	& \quad - \Big\langle  \frac{\partial \Re\ln \phi(a_{\rm real})}{\partial \theta^*} - i\frac{\partial \Im \ln \phi(a_{\rm repr})}{\partial \theta^*} \Big\rangle_{a_{\rm real} } \nonumber \\
    &\quad \cdot \Big\langle E_{\rm loc}(a_{\rm symm}) \Big\rangle_{a_{\rm real} },
\end{align}

where
\begin{equation}
    \langle \hat{O} \rangle_{a_{\rm real}} \equiv \frac{\sum_{a_{\rm real}}|\phi(a_{\rm real})|^{2} O(a_{\rm symm})}{\sum_{a_{\rm real}}|\phi(a_{\rm real})|^2}.
\end{equation}

The two rules of Eqs.~\eqref{amp_psi} and~\eqref{arg_psi} can be fused into a single effective amplitude, which takes its modulus from the sampled configuration and its phase from the representative. Defining
\begin{equation}
    \ln\widetilde{\phi}(a_{\rm real}) \equiv \Re \ln \phi(a_{\rm real}) + \iu\Im \ln\phi(a_{\rm repr}),
    \label{eq:phitilde}
\end{equation}
Eq.~\eqref{egradient} can be simplified to
\begin{align}
	\nonumber
	\frac{\partial E}{\partial \theta^*} = &\Big\langle \frac{\partial \ln\widetilde{\phi}^*(a_{\rm real})}{\partial \theta^*} E_{\rm loc}(a_{\rm symm}) \Big\rangle_{a_{\rm real} }\\
	&- \Big\langle \frac{\partial \ln \widetilde{\phi}^*(a_{\rm real})}{\partial \theta^*}\Big\rangle_{a_{\rm real} } \Big\langle E_{\rm loc}(a_{\rm symm}) \Big\rangle_{a_{\rm real} },
\end{align}
which makes explicit how the two bases divide the labor: configurations are drawn in real space from the generator, the variational derivatives act on the generator itself, evaluated at the sampled configuration $a_{\rm real}$ and at its representative $a_{\rm repr}$, and the Monte Carlo average $\langle\cdot\rangle_{a_{\rm real}}$ is taken over real-space samples; only the local energy $E_{\rm loc}(a_{\rm symm})$ is evaluated in the symmetric basis. Further details of the derivation are given in Appendix~\ref{app:energy}.

Note that a sampled $a_{\rm real}$ contributes only when its symmetric partner has nonzero norm, $N_{a_{\rm repr}} \neq 0$; this fails when the symmetry constraints of the chosen sector annihilate the equivalence class, and such samples are discarded.

\subsection{Quantum Circuits}
Designing variational ans\"atze within variational quantum circuits (VQCs) for simulating the eigenstates of many-body Hamiltonians remains a formidable challenge, as no universal strategy exists for constructing optimal circuit architectures. Each circuit must be carefully adapted to the specific Hamiltonian's interactions, symmetries, and entanglement structure~\cite{PeruzzoA2014,McCleanJR2016,KardashinA2020,ShangZX2023}. Moreover, designing quantum circuits that faithfully span the most relevant subspace is at least as difficult--if not even more so--than constructing circuits for individual eigenstates. Such circuits typically require deep architectures with extensive parameterization and limited transformation flexibility, which drastically increases the cost and complexity of optimization.

To alleviate these challenges and effectively reduce circuit depth, we draw inspiration from the density matrix renormalization group (DMRG) approach~\cite{WhiteSR1992}. In DMRG, the ground-state wave function is iteratively constructed as an entangled superposition of product eigenbases for system and environment blocks, where the most significant correlations are captured by retaining the eigenvectors of the RDM with the largest eigenvalues. Motivated by this principle, we replace the quantum circuits in the generator of Eq.~\eqref{phia} with an isometry formed from the $\chi$ dominant RDM eigenvectors of block size $b$. The RDM is extracted from a larger system whose ground state can be obtained via ED or DMRG. Additional shallow-depth quantum circuits may be applied on top of the isometry if necessary, as illustrated in Fig.~\ref{generator}(b). All results reported below use this isometric realization of $C_i$. In either realization $C_i$ enters Eq.~\eqref{phia} as the same kind of object---an isometry embedding the $\chi$ retained states into the $2^b$-dimensional block Hilbert space---so the sampling and gradient machinery developed above applies unchanged to both.

This substitution embeds the essential entanglement structure of the target Hamiltonian into a compact, low-rank representation, effectively reproducing DMRG's block-wise truncation mechanism while bypassing the need for costly variational optimization or hardware-specific gate synthesis. In this way, the method inherits DMRG's efficiency in identifying the most relevant subspace while maintaining compatibility with shallow-depth quantum circuit architectures. Formally, the relevant subspace is defined by the $\chi$ dominant eigenvectors of the reduced density matrix $\rho_b = \mathrm{Tr}_{\bar{b}} |\psi_{\rm ED}\rangle\langle\psi_{\rm ED}|$, where the partial trace is taken over all degrees of freedom outside block $b$. In practice, we expect this subspace to exhibit weak dependence on the total system size from which $\rho_b$ is obtained, since local entanglement patterns are governed primarily by short-range correlations. Furthermore, due to translational symmetry, the relevant subspace associated with $\rho_b$ remains nearly invariant under shifts of the block position along a chain of length $N$.

\section{Results}
\label{sec:results}
\begin{figure}
    \centering
    \includegraphics[width=\columnwidth]{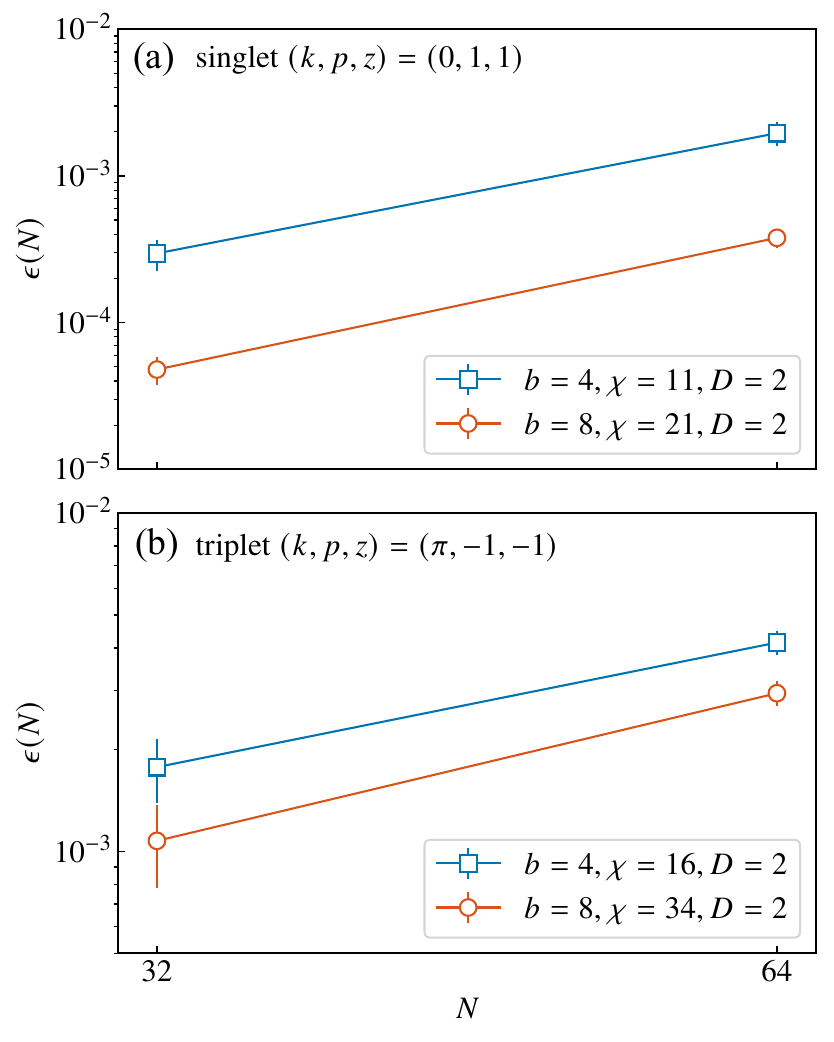}
    \caption{Relative energy error for the one-dimensional $J_1\!-\!J_2$ periodic chain, defined as $\epsilon(N) \equiv |[E(N) - E_{\rm DMRG}]/E_{\rm DMRG}|$, where $E(N)$ and $E_{\rm DMRG}$ are total energies, evaluated for chains of length $N = 32$ and 64 using MPS characterized by block size $b$, bond dimension $D$, and renormalized block-basis dimension $\chi$. (a) Singlet ground state with quantum numbers $(k,p,z) = (0,1,1)$ at $g = 0.2$. (b) Lowest triplet excitation with quantum numbers $(k,p,z) = (\pi,-1,-1)$ at $g = 0.25$.}
    \label{fig:energy}
\end{figure}

The Hamiltonian we use for demonstration is the frustrated spin-$1/2$ $J_1\!-\!J_2$ antiferromagnetic Heisenberg model on the 1D chain and 2D square lattice with periodic boundary conditions, formulated as
\begin{equation}
	\mathcal{H} = J_1 \sum_{\langle i, j \rangle} {\bm S}_i \cdot {\bm S}_j + J_2 \sum_{\langle \langle i, j \rangle \rangle} {\bm S}_i \cdot {\bm S}_j,
  	\label{hamilt}
\end{equation}
where the summations are performed over the nearest-neighbor pairs $\langle i, j \rangle$ and the next-nearest-neighbor pairs $\langle \langle i, j \rangle \rangle$ respectively. Throughout this paper, we set $J_1 = 1$ and denote $g = J_2 / J_1$.

The ground state phase diagrams of the 1D and 2D models reveal distinct frustration-driven behaviors. In 1D, the system passes from a Luttinger liquid (LL) for $g < g_{c1} \approx 0.2411$, to a valence bond solid (VBS) for $g_{c1} < g < g_{c2} \approx 0.5294$, and to an incommensurate spiral for $g > g_{c2}$~\cite{OkamotoK1992,BursillR1995,EggertS1996,WhiteSR1996,BrehmerS1998}. In 2D, competing interactions induce maximum frustration at $g=0.5$, manifesting as a quantum spin liquid (QSL)~\cite{SchulzH1996,CapriottiL2003,HuWJ2013,GongSS2014, PhysRevLett.121.107202, LiuWY2018}. We benchmark our method in the LL phase ($g=0.2$) and close to $g_{c1}$ ($g=0.25$) in 1D, and in the QSL phase ($g=0.5$) in 2D, against DMRG and ED results, respectively.

For the 1D Hamiltonian, we exploit translational symmetry $\mathcal{T}$, bond-mirror reflection $\mathcal{P}$, spin inversion $\mathcal{Z} = \prod_i \sigma^x_i$, and conservation of the total magnetization $S^z = \sum_i S^z_i$ to probe the ground state and lowest triplet excitation~\cite{BaoST2025,WangL2022}. For the 2D Hamiltonian, we employ translational symmetries $\mathcal{T}_x$, $\mathcal{T}_y$; bond-mirror reflections $\mathcal{P}_x$, $\mathcal{P}_y$; diagonal mirrors $\sigma_1$, $\sigma_2$; spin inversion $\mathcal{Z}$; and $S^z$ conservation for the ground state only~\cite{YangJ2022,WangL2022}.

\begin{figure*}
    \centering
    \includegraphics[width=\textwidth]{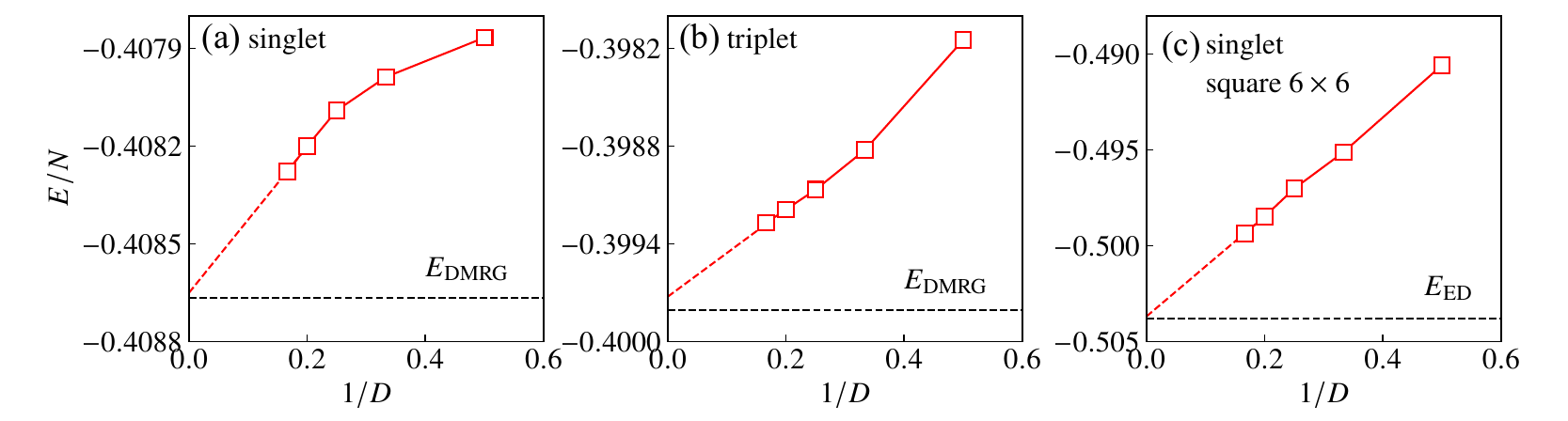}
    \caption{Variational energy per site versus inverse bond dimension. Results of the proposed eigensolver for a one-dimensional $J_1\!-\!J_2$ chain ($N = 64$) and a two-dimensional $6 \times 6$ torus, obtained using MPS with block size $b$, bond dimension $D$, and renormalized block-basis dimension $\chi$. Linear energy extrapolations (red dashed lines) use the three largest $D$ values, while black dashed lines indicate reference energies from ED and DMRG. (a) Singlet state with quantum numbers $(k,p,z) = (0,1,1)$ at $g = 0.2$, using $b = 4$, $D = 2\!-\!6$, and $\chi = 11$. (b) Triplet state with $(k,p,z) = (\pi,-1,-1)$ at $g = 0.25$, using $b = 4$, $D = 2\!-\!6$, and $\chi = 16$. (c) Singlet state of a $6\times6$ lattice with $(k_x,k_y,p_x,p_y,\sigma_1,\sigma_2,z) = (0,0,1,1,1,1,1)$ at $g = 0.5$, using $b=2\times 2$, $D = 2\!-\!6$ and $\chi = 16$.}
    \label{fig:extrapolate}
\end{figure*}

We investigate the 1D $J_1\!-\!J_2$ model for system sizes $N=32$ and $64$ at $g=0.2$ and 0.25, using MPS with block sizes $b=4,8$, renormalized block-basis dimension $\chi$, and bond dimension $D$. As outlined in the method, the circuits $C_i$ are realized as the isometry built from the dominant RDM eigenvectors for block size $b$, computed from an exactly solvable $N=16$ chain via ED.

Fig.~\ref{fig:energy} displays the relative energy error, $\epsilon(N)=|[E(N)-E_{\rm DMRG}]/ E_{\rm DMRG}|$, of the variational energies for the singlet ground state $(k,p,z)=(0,1,1)$ at $g=0.2$ and the lowest triplet excitation $(k,p,z)=(\pi,-1,-1)$ at $g=0.25$ respectively. For the singlet, the renormalized block-basis dimension is set to $\chi=11$ for $b=4$ and $\chi=21$ for $b=8$, determined by the eigenvalue decay of the corresponding RDM $\rho_b={\rm Tr}_{\bar b}|\psi_{\rm ED}(k,p,z)\rangle\langle\psi_{\rm ED}(k,p,z)|$, where ${\rm Tr}_{\bar b}$ denotes the trace over all spins outside block $b$, and $|\psi_{\rm ED}(k,p,z)\rangle$ denotes the desired $(k,p,z)$ eigenstate of a chain of size $N=16$ obtained by ED. Similarly, for the triplet, the renormalized block-basis dimension is set to $\chi=16$ for $b=4$ and $\chi=34$ for $b=8$. The number of variational parameters equals $2\chi D+(N/b-2)\chi D^2$.

Fig.~\ref{fig:energy}(a) shows that, for the singlet at $N=32$, energy precisions of $3\times 10^{-4}$ and $5\times 10^{-5}$ are achieved with block sizes $b=4$ (308 parameters) and $b=8$ (252 parameters), respectively. For $N=64$, precisions of $2\times 10^{-3}$ and $4\times 10^{-4}$ are obtained with $b=4$ (660 parameters) and $b=8$ (588 parameters), respectively. We observe that larger block sizes $b$ yield an order of magnitude higher precision for a similar number of variational parameters. However, for the same block size $b$, the $N=64$ chain has an order of magnitude lower precision than $N=32$.

Fig.~\ref{fig:energy}(b) shows that, for the triplet at $N=32$, energy precisions of $1.8\times 10^{-3}$ and $1\times 10^{-3}$ are achieved with block sizes $b=4$ (448 parameters) and $b=8$ (408 parameters), respectively. For $N=64$, precisions of $4\times 10^{-3}$ and $3\times 10^{-3}$ are obtained with $b=4$ (960 parameters) and $b=8$ (952 parameters), respectively. Compared to the singlet, increasing $b$ yields less significant precision improvement.

To enhance the accuracy of the proposed eigensolver in regimes with strong entanglement, we perform linear extrapolations of the variational energy with respect to the inverse bond dimension 1/$D$ for both the one and two dimensional $J_1\!-\!J_2$ Heisenberg models. For the 64-site periodic chain at $g = 0.2, 0.25$ and the $6 \times 6$ square lattice at $g = 0.5$, we fix the block size to $b = 4$ in one dimension and $b = 2\times 2$ in two dimensions, where the corresponding reduced density matrix $\rho_b$ is obtained from an exact diagonalization wavefunction on a $4\times 4$ torus. The bond dimension $D$ is then varied to assess convergence. Figure~\ref{fig:extrapolate} displays the variational energy per site as a function of $1/D$ on a linear scale. In both systems, simple linear extrapolation using the three largest $D$ values yields an absolute error in the energy per site on the order of $10^{-5}$ for the singlet ground state in one dimension, $10^{-4}$ for the triplet sector, and $10^{-5}$ for the $6\times6$ torus. Despite employing extremely small bond dimensions ($D \le 6$), the extrapolated energies show excellent agreement with high-accuracy benchmarks. This is the mechanism anticipated above: the symmetry-adapted basis itself carries the long-range entanglement, leaving the MPS backbone to represent only what remains. Appendix~\ref{app:ee} makes the effect quantitative---at $D=1$ the symmetric state already reproduces the bipartite entanglement entropy of the exact ground state ($S_A = 1.2895$ against $1.2832$), whereas the same generator used directly as a real-space wave function is a product state across the partition. These results indicate that the method attains DMRG-level precision at these system sizes with a dramatically reduced number of variational parameters.

For two-dimensional quantum lattice models, projected entangled pair states (PEPS) provide a more natural tensor-network geometry than an MPS representation, as their connectivity directly reflects the spatial structure of the underlying lattice~\cite{VerstraeteCirac2004PEPS, Cirac2021}. In the present work, we employ an MPS as the tensor-network bridge primarily for simplicity and to demonstrate the core concept of combining local quantum circuits with a global tensor-network structure. Nonetheless, the methodology developed here is not restricted to an MPS backbone. The symmetry-adapted mapping, the isometric embedding, and the direct real-space sampling of the generator can in principle be extended to PEPS structures. In this sense, PEPS represents a more suitable choice for large-scale two-dimensional applications of the proposed framework. Developing such a PEPS-based formulation of our tensor-network-bridged quantum-circuit method is thus an important direction for future research.

\section{Conclusions}
\label{sec:conclusions}
We have introduced a symmetry-adapted eigensolver that separates where the wave function is sampled from where it is defined. Parallel quantum circuits capture the entanglement within each block while an MPS backbone bridges the blocks, but the amplitude this composite object generates serves only to supply real-space configurations; the variational state itself is defined in a symmetry-adapted basis by two separate rules, taking its modulus from the total weight of an entire equivalence class and its phase from the representative configuration alone. Because the configurations within a class are combined in probability rather than in amplitude, the result is not a group projection: intra-class phases never interfere. The resulting many-to-one correspondence both compresses the relevant Hilbert space and enhances bipartite entanglement across real-space partitions, as quantified in Appendix~\ref{app:ee}.

The framework further exploits isometric maps derived from reduced density matrices (RDM) to identify and retain the most relevant subspaces of the Hilbert space. These RDM-based isometries alleviate the need for deep and highly parameterized quantum circuits, substantially easing the burden of circuit design and variational optimization. Because the relevant subspace obtained from local RDM depends weakly on the overall system size and position along translationally symmetric chains, the scheme provides a stable and transferable foundation for treating larger systems.

Benchmark calculations on both one and two dimensional $J_1\!-\!J_2$ antiferromagnetic Heisenberg models validate the power of this approach. Remarkably, even with extremely small bond dimensions ($D \le 6$), linear extrapolation of the variational energy yields absolute errors in the energy per site as low as $10^{-5}$--comparable to conventional density matrix renormalization group (DMRG) calculations employing bond dimensions in the thousands. This accuracy is attributable to the symmetry-adapted basis rather than to the tensor network: the basis carries the long-range entanglement, leaving the backbone to represent the remainder within a dramatically reduced parameter space. Specifically, the cost per Monte Carlo sample separates into two parts. Generating a configuration requires the conditional reduced density matrices of Fig.~\ref{sampling}, whose cost is that of standard MPS operations, $\mathcal{O}(n_b \chi^2 D^3)$ in time with a memory footprint $\mathcal{O}(n_b \chi D^2)$. Evaluating the local energy is the more demanding step: the Hamiltonian connects each sampled state to $\mathcal{O}(N)$ others, and the modulus of every connected symmetric amplitude requires the weights of all $|\mathcal{G}|$ members of its equivalence class, so this step carries an additional factor $N|\mathcal{G}|$, itself growing with system size since $\mathcal{G}$ contains the $N$ lattice translations. The symmetry aggregation is therefore the dominant classical cost. It remains a favorable trade: reaching a given accuracy at $D \le 6$ rather than at the far larger bond dimensions required by conventional DMRG reduces the $D^3$ term by several orders of magnitude, outweighing the $N|\mathcal{G}|$ overhead, with a corresponding reduction in memory footprint.

In addition, the DMRG or tensor-network obtained block isometries and local tensors can serve as natural initializations for our hybrid optimizations, especially in the large-$b$ and large-$D$ regime, thereby accelerating convergence and improving computational efficiency. Together, these features establish a coherent framework in which real-space quantum sampling, symmetry-constrained measurement, and tensor-network reconstruction operate synergistically.

Finally, we discuss a potential implementation of the
proposed framework on quantum hardware. Two features of the
ansatz are well matched to a distributed architecture. First, the blocks are
bridged by the classical tensor network rather than by quantum links, so the
individual processors never need to share entanglement; the most demanding
requirement of distributed quantum computing therefore does not arise here.
Second, the generative step is directly executable. Since Eq.~\eqref{phia}
is a matrix product state of bond dimension $D$ whose physical leg is dilated
by the isometry $C_i$, the state $|\phi\rangle$ can be prepared
sequentially~\cite{Schon2005,Ran2020}: a bond register of
$\lceil \log_2 D\rceil$ qubits is held coherently while each block of $b$
qubits is generated by a unitary embedding of $B_i$ followed by $C_i$,
measured in the computational basis, and reset for the next
block~\cite{Chertkov2022,Foss-Feig2022}. The recorded bit string is then an
exact sample of $p(a_{\rm real})=|\phi(a_{\rm real})|^2$, obtained without
amplitude estimation or tomography, and the number of simultaneously live
qubits is $\lceil\log_2 D\rceil + b$, independent of the system size $N$.

The energy estimator, by contrast, does not
transfer as directly. Equations~\eqref{amp_psi} and~\eqref{arg_psi} require
the modulus $|\phi|$ at every configuration of an equivalence class and the
phase $\arg\phi(a_{\rm repr})$, while $E_{\rm loc}$ requires the ratios
$\psi(a'_{\rm symm})/\psi(a_{\rm symm})$. These are amplitude queries rather
than expectation values, and their cost on a quantum processor grows both
with the block size $b$ and with the order of the symmetry group. Recasting
the estimator in terms of directly measurable expectation values is thus the
central step toward a hardware realization, and we leave it to future work.

Overall, this work shows that pairing a real-space generator with a symmetry-adapted variational basis relieves much of the entanglement burden that limits conventional tensor-network ansatzes. As demonstrated here the eigensolver is entirely classical, and its accuracy at very small bond dimension makes it attractive in that setting on its own. At the same time the architecture is well matched to distributed quantum hardware, since the blocks are bridged classically and the generative step is directly executable; turning it into a genuine quantum-classical algorithm hinges on recasting the estimator in terms of measurable expectation values.

\begin{acknowledgments}
We would like to thank P. Zhang for useful discussion. This work was supported by the National Natural Science Foundation of China (Grant No.~12374150).
\end{acknowledgments}
\bibliography{main}

\clearpage
\appendix
\onecolumngrid
\section{Energy Derivation}
\label{app:energy}
The Wirtinger derivatives~\cite{Kreutz-DelgadoK2009} are defined as

\begin{subequations}
    \begin{align}
        \frac{\partial}{\partial z} & \equiv \frac{1}{2}\left(\frac{\partial}{\partial x}-\iu\frac{\partial}{\partial y}\right), \\
        \frac{\partial}{\partial z^{*}} &\equiv\frac{1}{2}\left(\frac{\partial}{\partial x}+\iu\frac{\partial}{\partial y}\right),
    \end{align}
\end{subequations}
where $z = x+\iu y$, with $x,y \in \mathbb{R}$. For a real-valued loss function $L$ the direction of steepest ascent in the $(x,y)$ plane is $2\,\partial L/\partial z^{*}$, so a descent step of size $\alpha/2$ in the real variables takes the compact form $z^{(n+1)} = z^{(n)} - \alpha\,\partial L / \partial z^{*}$; derivatives are therefore taken throughout with respect to the conjugated variables. Writing $\theta$ for any of the complex variational parameters --- the entries of the MPS tensors $B_i$, and more generally the parameters of the block circuits $C_i$ --- and noting that the energy $E$ is real, the quotient rule gives

\begin{align}
    \label{eq:PXE}
    \frac{\partial E}{\partial\theta^{*}}&=\frac{\partial}{\partial\theta^{*}}\frac{\langle\psi\vert \mathcal{H} \vert\psi\rangle}{\langle\psi\vert\psi\rangle}\nonumber \\
    &=\frac{1}{\langle\psi\vert\psi\rangle}\frac{\partial}{\partial\theta^{*}}\langle\psi\vert \mathcal{H} \vert\psi\rangle-\frac{1}{\langle\psi\vert\psi\rangle}\langle\psi\vert \mathcal{H} \vert\psi\rangle\frac{1}{\langle\psi\vert\psi\rangle}\frac{\partial}{\partial\theta^{*}}\langle\psi\vert\psi\rangle
\end{align}

Two properties of the ansatz control the manipulations that follow. First, the generator $\phi(a_{\rm real})$ of Eq.~\eqref{phia} is a contraction in which no tensor entry is ever conjugated, so $\phi$ is a holomorphic function of the parameters,
\begin{equation}
    \frac{\partial\phi(a_{\rm real})}{\partial\theta^{*}}=0 ,
    \label{eq:holo}
\end{equation}
and hence $\partial\ln\phi/\partial\theta^{*}=0$ as well. Writing $\Re\ln\phi=\frac{1}{2}\left(\ln\phi+\ln\phi^{*}\right)$ and $\Im\ln\phi=\frac{1}{2\iu}\left(\ln\phi-\ln\phi^{*}\right)$, Eq.~\eqref{eq:holo} yields the two identities used repeatedly below,
\begin{subequations}
    \begin{align}
        \frac{\partial\,\Im\ln\phi}{\partial\theta^{*}} &= \iu\,\frac{\partial\,\Re\ln\phi}{\partial\theta^{*}}, \label{eq:idA}\\
        \frac{\partial\left|\phi\right|^{2}}{\partial\theta^{*}} &= 2\left|\phi\right|^{2}\,\frac{\partial\,\Re\ln\phi}{\partial\theta^{*}} . \label{eq:idB}
    \end{align}
\end{subequations}
Equation~\eqref{eq:idB} follows from $\partial_{\theta^{*}}(\phi^{*}\phi)=\phi\,\partial_{\theta^{*}}\phi^{*}+\phi^{*}\partial_{\theta^{*}}\phi$, whose second term is annihilated by Eq.~\eqref{eq:holo}; it is the origin of the factor of two in Eq.~\eqref{eq:PXpsi} below. Note that $\left|\phi\right|^{2}$ is of course \emph{not} holomorphic, nor is the energy $E$, which is real: holomorphy is a property of the amplitude alone, and it is precisely what makes one of the two terms in each product rule vanish.

Second, and by contrast, the symmetric amplitude $\psi(a_{\rm symm})$ is \emph{not} holomorphic. Its modulus is drawn from an entire equivalence class while its phase is taken from the representative alone, and the two do not in general agree. Applying Eqs.~\eqref{eq:idA} and~\eqref{eq:idB} to Eqs.~\eqref{amp_psi} and~\eqref{arg_psi} gives
\begin{equation}
    \frac{\partial\ln\psi\left(a_{\rm symm}\right)}{\partial\theta^{*}}
    =\Big\langle \frac{\partial\,\Re\ln\phi\left(a_{\rm real}\right)}{\partial\theta^{*}}\Big\rangle_{\{g a_{\rm repr}\}}
    -\frac{\partial\,\Re\ln\phi\left(a_{\rm repr}\right)}{\partial\theta^{*}},
    \label{eq:nonholo}
\end{equation}
where $\langle\cdot\rangle_{\{g a_{\rm repr}\}}$ is the average over the equivalence class weighted by $\left|\phi(a_{\rm real})\right|^{2}$. The right-hand side vanishes only when the class-averaged logarithmic derivative coincides with its value at the representative, that is, when the generator is already symmetric within the class; the departure from holomorphy thus measures how far $\phi$ is from being symmetric. Because $\partial\ln\psi/\partial\theta^{*}\neq0$, both terms must be retained in Eq.~\eqref{eq:PXH}, and the gradient cannot be expressed through $\psi$ alone. 

Using the identity $\sum_{a_{\rm repr}}\sum_{a_{\rm real}\in\{g a_{\rm repr}\}}=\sum_{a_{\rm real}}$, the derivative of the norm becomes

\begin{align}
    \label{eq:PXpsi}
    \frac{\partial}{\partial\theta^{*}}\langle\psi\vert\psi\rangle&=\frac{\partial}{\partial\theta^{*}}\sum_{a_{\rm symm}}\psi^{*}\left(a_{\rm symm}\right)\psi\left(a_{\rm symm}\right) \nonumber \\
    &=\frac{\partial}{\partial\theta^{*}}\sum_{a_{\rm real}}\phi^{*}\left(a_{\rm real}\right)\phi\left(a_{\rm real}\right)\nonumber \\
    &=2\sum_{a_{\rm real}}\frac{\partial\Re\ln\phi\left(a_{\rm real}\right)}{\partial\theta^{*}}\left|\phi\left(a_{\rm real}\right)\right|^{2}.
\end{align}

\begin{align}
    \label{eq:PXH}
    \frac{\partial}{\partial\theta^{*}}\langle\psi\vert \mathcal{H} \vert\psi\rangle&=\sum_{a_{\rm symm}}\frac{\partial\psi^{*}\left(a_{\rm symm}\right)}{\partial\theta^{*}}\sum_{a'_{\rm symm}}\psi\left(a'_{\rm symm}\right)\mathcal{H}_{a_{\rm symm},a'_{\rm symm}} + \sum_{a_{\rm symm}}\psi^{*}\left(a_{\rm symm}\right)\sum_{a'_{\rm symm}}\frac{\partial\psi\left(a'_{\rm symm}\right)}{\partial\theta^{*}}\mathcal{H}_{a_{\rm symm},a'_{\rm symm}} \nonumber \\
    &=\sum_{a_{\rm symm}}\left|\psi\left(a_{\rm symm}\right)\right|^{2} \left[ \frac{\partial\ln\psi^{*}\left(a_{\rm symm}\right)}{\partial\theta^{*}}E_{{\rm loc}}\left(a_{\rm symm}\right) + \frac{\partial\ln\psi\left(a_{\rm symm}\right)}{\partial\theta^{*}}E_{{\rm loc}}^{*}\left(a_{\rm symm}\right) \right] \nonumber \\
    &=2\sum_{a_{\rm symm}}\left|\psi\left(a_{\rm symm}\right)\right|^{2} \left[ \frac{\partial\Re\ln\psi\left(a_{\rm symm}\right)}{\partial\theta^{*}}\Re E_{{\rm loc}}\left(a_{\rm symm}\right) + \frac{\partial\Im\ln\psi\left(a_{\rm symm}\right)}{\partial\theta^{*}}\Im E_{{\rm loc}}\left(a_{\rm symm}\right) \right] \nonumber\\
    &=2\sum_{a_{\rm real}}\left|\phi\left(a_{\rm real}\right)\right|^{2}\left[ \frac{\partial\Re\ln\phi\left(a_{\rm real}\right)}{\partial\theta^{*}}\Re E_{{\rm loc}}\left(a_{\rm real}\right) + \frac{\partial\Im\ln\phi\left(a_{\rm repr}\right)}{\partial\theta^{*}}\Im E_{{\rm loc}}\left(a_{\rm real}\right) \right],
\end{align}
where $E_{{\rm loc}}\left(a_{\rm symm}\right) = \sum_{a'_{\rm symm}} \mathcal{H}_{a_{\rm symm},a'_{\rm symm}} [\psi\left(a'_{\rm symm}\right)/\psi\left(a_{\rm symm}\right)]$. Here $a_{\rm real}$ belongs to the equivalence class mapped to the symmetric state $a_{\rm symm}$, and $a_{\rm repr}$ is its representative state.

\begin{subequations}
    \label{eq:PXln}
    \begin{align}
        \frac{\partial\Re\ln\psi\left(a_{\rm symm}\right)}{\partial\theta^{*}} &= \frac{1}{2}\frac{1}{\sum_{\left\{ ga_{\rm repr}\right\} }\left|\phi\left(ga_{\rm repr}\right)\right|^{2}}\frac{\partial}{\partial\theta^{*}}\sum_{\left\{ ga_{\rm repr}\right\} }\left|\phi\left(ga_{\rm repr}\right)\right|^{2}\nonumber\\
        &= \frac{1}{\left|\psi\left(a_{\rm symm}\right)\right|^{2}}\sum_{\left\{ ga_{\rm repr}\right\} }\left|\phi\left(ga_{\rm repr}\right)\right|^{2}\frac{\partial\Re\ln\phi\left(ga_{\rm repr}\right)}{\partial\theta^{*}}, \\
        \frac{\partial\Im\ln\psi\left(a_{\rm symm}\right)}{\partial\theta^{*}} &= \frac{\partial\Im\ln\phi\left(a_{\rm repr}\right)}{\partial\theta^{*}}.
    \end{align}
\end{subequations}

Substituting Eqs.~\eqref{eq:PXpsi}, \eqref{eq:PXln}, and~\eqref{eq:PXH} into Eq.~\eqref{eq:PXE} yields
\begin{align}
    \frac{\partial E}{\partial\theta^{*}} & = 2\left\langle \frac{\partial\Re\ln\phi\left(a_{\rm real}\right)}{\partial\theta^*}\Re E_{{\rm loc}}\left(a_{\rm real}\right)\right\rangle_{a_{\rm real}}-2\left\langle E_{{\rm loc}}\right\rangle _{a_{\rm real}}\left\langle \frac{\partial \Re\ln\phi\left(a_{\rm real}\right)}{\partial\theta^{*}}\right \rangle_{a_{\rm real}}  +2\left\langle \frac{\partial\Im\ln\phi\left(a_{\rm repr}\right)}{\partial\theta^{*}}\Im E_{{\rm loc}}\left(a_{\rm real}\right)\right\rangle _{a_{\rm real}}.
\end{align}


\section{Symmetric States}
\label{app:sps}
Following the derivation in Ref.~\cite{SandvikAW2010_lecture}, we provide a detailed derivation of the symmetric states, their normalization factors, and the conversion between their symmetric and real-space representations. We then discuss the allowed quantum numbers for the symmetry group considered in our work.

\subsection{General Formalism}

Let $\mathcal{G}$ denote the symmetry group and let $\chi_g$ be the phase factor associated with each element $g\in\mathcal{G}$, satisfying $\chi_{g_1g_2}=\chi_{g_1}\chi_{g_2}$ and $\chi_{g^{-1}}=\chi_g^*$.

The corresponding symmetric basis state is
\begin{equation}
| a_{\text{symm}}\rangle = \frac{1}{\sqrt{N_{a_{\text{repr}}}}}\sum_{g\in \mathcal{G}} \chi_g^* g | a_{\text{repr}}\rangle,
\end{equation}
The action of any group element $g$ then follows as
\begin{align}
g | a_{\text{symm}}\rangle &= \frac{1}{\sqrt{N_{a_{\text{repr}}}}}\sum_{g_{1}\in \mathcal{G}}\chi_{g_{1}}^* g g_{1} | a_{\text{repr}}\rangle \nonumber \\
 &= \frac{1}{\sqrt{N_{a_{\text{repr}}}}} \chi_{g} \sum_{g_{2}\in \mathcal{G}} \chi_{g_{2}}^* g_{2} | a_{\text{repr}}\rangle \nonumber \\
 &= \chi_g | a_{\text{symm}}\rangle.
\end{align}

To expand this state in the real-space basis, let $\{g a_{\text{repr}}\}=\{g|a_{\text{repr}}\rangle\mid g\in\mathcal{G}\}$ denote the orbit (equivalence class) of $|a_{\text{repr}}\rangle$. For each configuration $|a_{\text{real}}\rangle\in\{g a_{\text{repr}}\}$, define the subset of group elements that map $|a_{\text{repr}}\rangle$ to $|a_{\text{real}}\rangle$ by
\begin{equation}
G_{a_{\text{real}}} = \{ g \in \mathcal{G} \mid g |a_{\text{repr}}\rangle = |a_{\text{real}}\rangle \}.
\end{equation}
Partitioning the sum over group elements according to their action on $|a_{\text{repr}}\rangle$ gives
\begin{align}
| a_{\text{symm}}\rangle &= \frac{1}{\sqrt{N_{a_{\text{repr}}}}} \sum_{a_{\text{real}} \in \{ g a_{\text{repr}} \}} \sum_{g \in G_{a_{\text{real}}}} \chi_g^* | a_{\text{real}}\rangle \nonumber \\
 &= \frac{1}{\sqrt{N_{a_{\text{repr}}}}} \sum_{a_{\text{real}} \in \{g a_{\text{repr}}\}} \Lambda_{a_{\text{real}}} | a_{\text{real}}\rangle,
\end{align}
where the phase sum over the mapping subset is
\begin{equation}
\Lambda_{a_{\text{real}}} = \sum_{g \in G_{a_{\text{real}}}} \chi_g^*.
\end{equation}
The normalization factor follows from the norm of the unnormalized group sum,
\begin{align}
N_{a_{\text{repr}}} &= \sum_{g_2 \in \mathcal{G}} \sum_{g_1 \in \mathcal{G}} \chi_{g_2} \chi_{g_1}^* \langle a_{\text{repr}} | g_2^{-1} g_1 | a_{\text{repr}} \rangle \nonumber \\
 &= |\mathcal{G}| \sum_{g \in \mathcal{G}} \chi_g^* \langle a_{\text{repr}} | g | a_{\text{repr}} \rangle \nonumber \\
 &= |\mathcal{G}| \sum_{g \in G_{a_{\text{repr}}}} \chi_g^*.
\end{align}
Because $G_{a_{\text{repr}}}$ is the stabilizer subgroup of $|a_{\text{repr}}\rangle$, $\sum_{g\in G_{a_{\text{repr}}}}\chi_g^*=\Lambda_{a_{\text{repr}}}$. Since $\Lambda_{a_{\text{repr}}}$ is real, the normalization factor is
\begin{equation}
N_{a_{\text{repr}}} = |\mathcal{G}| \Lambda_{a_{\text{repr}}}.
\end{equation}
The full wave function can then be expanded in the symmetric basis and rewritten in the real-space basis as
\begin{align}
| \psi \rangle &= \sum_{a_{\text{symm}}} \psi(a_{\text{symm}}) | a_{\text{symm}}\rangle \nonumber \\
 &= \sum_{a_{\text{repr}}} \psi(a_{\text{symm}}) \frac{1}{\sqrt{N_{a_{\text{repr}}}}} \sum_{a_{\text{real}} \in \{ g a_{\text{repr}} \}} \Lambda_{a_{\text{real}}} | a_{\text{real}}\rangle \nonumber \\
 &= \sum_{a_{\text{real}}} \psi(a_{\text{symm}}) \frac{\Lambda_{a_{\text{real}}}}{\sqrt{N_{a_{\text{repr}}}}} | a_{\text{real}}\rangle \nonumber \\
 &= \sum_{a_{\text{real}}} \psi(a_{\text{real}}) | a_{\text{real}}\rangle.
\end{align}
Comparing coefficients gives
\begin{equation}
    \label{eq:psireal}
\psi(a_{\text{real}}) = \psi(a_{\text{symm}}) \frac{\Lambda_{a_{\text{real}}}}{\sqrt{N_{a_{\text{repr}}}}}.
\end{equation}

\subsection{Concrete Example}

For the specific 1D case in our paper, consider a symmetry group $\mathcal{G}$ consisting of the translational ($\mathcal{T}$), bond-mirror reflection ($\mathcal{P}$), and spin-inversion ($\mathcal{Z}$) operators. A symmetric state under this group can be characterized by quantum numbers $k, p, z$ and written as:

\begin{align}
    \vert a_{{\rm symm}}\left(k,p,z\right)\rangle & =\frac{1}{\sqrt{N_{a_{\rm repr}}}}\sum_{r=0}^{N-1}\sum_{u=0}^{1}\sum_{v=0}^{1}e^{-\iu k\cdot r}\mathcal{T}^{r}\left(p \mathcal{P}\right)^{u}\left(z \mathcal{Z}\right)^{v}\vert a_{{\rm repr}}\rangle\nonumber \\
    & =\frac{1}{\sqrt{N_{a_{\rm repr}}}}\sum_{r=0}^{N-1}e^{-\iu k\cdot r}\mathcal{T}^{r}\left(1+p\mathcal{P}\right)\left(1+z\mathcal{Z}\right)\vert a_{{\rm repr}}\rangle,
\end{align}

where $\vert a_{{\rm repr}}\rangle$ is the unique representative state, chosen as the lexicographically smallest configuration in the orbit $\{g\vert a_{{\rm real}}\rangle\mid g\in\mathcal{G}\}$, and $N_{a_{\rm repr}}$ is the normalization factor.

To determine $N_{a_{\rm repr}}$, we find all $\left\{ r_{i},u_{i},v_{i}\right\}$ that satisfy $\mathcal{T}^{r_{i}}\mathcal{P}^{u_{i}}\mathcal{Z}^{v_{i}}\vert a_{{\rm repr}}\rangle=\vert a_{{\rm repr}}\rangle$:

\begin{align}
    N_{a_{\rm repr}} & =\sum_{r,r'=0}^{N-1}\sum_{u,u'=0}^{1}\sum_{v,v'=0}^{1}\langle a_{{\rm repr}}\vert e^{-\iu k\cdot\left(r-r'\right)}\mathcal{T}^{r-r'}\left(p\mathcal{P}\right)^{u-u'}\left(z\mathcal{Z}\right)^{v-v'}\vert a_{{\rm repr}}\rangle\nonumber \\
    & =|\mathcal{G}|\sum_{r=0}^{N-1}\sum_{u=0}^{1}\sum_{v=0}^{1}\langle a_{{\rm repr}}\vert e^{-\iu k\cdot r}\mathcal{T}^{r}\left(p\mathcal{P}\right)^{u}\left(z\mathcal{Z}\right)^{v}\vert a_{{\rm repr}}\rangle\nonumber \\
    & =|\mathcal{G}|\sum_{i}e^{-\iu k\cdot r_{i}}p^{u_{i}}z^{v_{i}},
\end{align}
where $|\mathcal{G}|$ is the number of elements in the group $\mathcal{G}$. For the group consisting of $\mathcal{T}$, $\mathcal{P}$, and $\mathcal{Z}$, it takes the value of $4N$.

Next, we consider the allowed quantum numbers for the current set. The phase is determined by the target quantum numbers $(k,p,z)$ such that the symmetric basis states satisfy:

\begin{subequations}

\begin{align}
    \mathcal{T}\vert a_{{\rm symm}}\left(k,p,z\right)\rangle & =e^{\iu k}\vert a_{{\rm symm}}\left(k,p,z\right)\rangle,\\
    \mathcal{P}\vert a_{{\rm symm}}\left(k,p,z\right)\rangle & =p\vert a_{{\rm symm}}\left(k,p,z\right)\rangle,\\
    \mathcal{Z}\vert a_{{\rm symm}}\left(k,p,z\right)\rangle & =z\vert a_{{\rm symm}}\left(k,p,z\right)\rangle.
\end{align}

\end{subequations}

For $\mathcal{T}$,

\begin{align}
    \mathcal{T}\vert a_{{\rm symm}}\left(k,p,z\right)\rangle & =\frac{e^{ik}}{\sqrt{N_{a_{\rm repr}}}}\sum_{r=0}^{N-1}\sum_{u=0}^{1}\sum_{v=0}^{1}e^{-\iu k\cdot\left(r+1\right)}e^{\iu k}\mathcal{T}^{r+1}\left(p\mathcal{P}\right)^{u}\left(z\mathcal{Z}\right)^{v}\vert a_{{\rm repr}}\rangle \nonumber\\
    & =e^{\iu k}\vert a_{{\rm symm}}\left(k,p,z\right)\rangle.
\end{align}

For the reflection operator $\mathcal{P}$, using the relations $\mathcal{P}\mathcal{T}=\mathcal{T}^{-1}\mathcal{P}$ and $\mathcal{Z}\mathcal{P}=\mathcal{P}\mathcal{Z}$, we have:

\begin{align}
    \mathcal{P}\vert a_{{\rm symm}}\left(k,p,z\right)\rangle & =\frac{1}{\sqrt{N_{a_{\rm repr}}}}\sum_{r=0}^{N-1}\sum_{u=0}^{1}\sum_{v=0}^{1}e^{-\iu k\cdot r}\mathcal{T}^{-r}\mathcal{P}\left(p\mathcal{P}\right)^{u}\left(z\mathcal{Z}\right)^{v}\vert a_{{\rm repr}}\rangle \nonumber \\
    & =p\frac{1}{\sqrt{N_{a_{\rm repr}}}}\sum_{r=0}^{N-1}\sum_{u=0}^{1}\sum_{v=0}^{1}e^{\iu k\cdot r}\mathcal{T}^{r}\left(p\mathcal{P}\right)^{u}\left(z\mathcal{Z}\right)^{v}\vert a_{{\rm repr}}\rangle,
\end{align}

which implies $e^{\iu k} = e^{-\iu k}$ (or equivalently, $e^{\iu k\cdot r}=e^{-\iu k\cdot r}$ for all $r$), yielding the allowed momentum values:

\begin{equation}
k=0\quad {\rm or} \quad \pi.
\end{equation}

For the spin-inversion operator $\mathcal{Z}$, which acts only in spin space while $\mathcal{T}$ and $\mathcal{P}$ act only on the lattice sites, the commutation relations $\left[\mathcal{T},\mathcal{Z}\right]=0$ and $\left[\mathcal{P},\mathcal{Z}\right]=0$ hold as operator identities. The constraint $\sum_i S^z_i=0$ plays a different role: $\mathcal{Z}$ maps the sector of total magnetization $m$ onto that of $-m$, so it leaves the sector invariant only at $m=0$. The remaining pair requires more care, since $\mathcal{P}\mathcal{T}=\mathcal{T}^{-1}\mathcal{P}$ and hence $\left[\mathcal{T},\mathcal{P}\right]\neq0$ in general; within a momentum eigenspace, however, $\mathcal{T}$ acts as the scalar $e^{\iu k}$, and $\mathcal{P}$ preserves that eigenspace precisely when $k=0$ or $\pi$, as found above. Taken together, in the sector $\sum_i S^z_i=0$ and at $k=0$ or $\pi$ the three operators $\mathcal{T}$, $\mathcal{P}$ and $\mathcal{Z}$ commute pairwise and can therefore be diagonalized simultaneously, so that $(k,p,z)$ forms a consistent set of quantum numbers.

To construct the Hamiltonian matrix in the symmetric basis, we separate its diagonal and off-diagonal two-site terms and use one-based periodic indexing, $S_0^{\alpha}\equiv S_N^{\alpha}$ and $S_{N+1}^{\alpha}\equiv S_1^{\alpha}$.

Taking the nearest-neighbor interaction as an example, we write $\mathcal{H}=\sum_{j=0}^{N}\mathcal{H}_{j}$:

\begin{subequations}
    \begin{align}
        \mathcal{H}_{0} & =J\sum_{i=1}^{N}S_{i}^{z}S_{i+1}^{z},\\
        \mathcal{H}_{j} & =\frac{J}{2}\left(S_{j-1}^{+}S_{j}^{-}+S_{j-1}^{-}S_{j}^{+}\right), & j=1,\cdots,N.
    \end{align}
\end{subequations}

The matrix element in the real-space basis is given by:

\begin{equation}
h_{j}\left(a_{{\rm repr}}\right)=\langle b_{{\rm real}}^{j}\vert \mathcal{H}_{j}\vert a_{{\rm repr}}\rangle.
\end{equation}
For the symmetric basis, we have:

\begin{align}
    \mathcal{H} \vert a_{{\rm symm}}\left(k,p,z\right)\rangle & =\frac{1}{\sqrt{N_{a_{\rm repr}}}}\sum_{r=0}^{N-1}\sum_{u=0}^{1}\sum_{v=0}^{1}e^{-\iu k\cdot r}\mathcal{T}^{r}\left(p\mathcal{P}\right)^{u}\left(z\mathcal{Z}\right)^{v}\mathcal{H}\vert a_{{\rm repr}}\rangle \nonumber \\
    & =\frac{1}{\sqrt{N_{a_{\rm repr}}}}\sum_{r=0}^{N-1}\sum_{u=0}^{1}\sum_{v=0}^{1}e^{-\iu k\cdot r}\mathcal{T}^{r}\left(p\mathcal{P}\right)^{u}\left(z\mathcal{Z}\right)^{v}\sum_{j}h_{j}\left(a_{{\rm repr}}\right)\vert b_{{\rm real}}^{j}\rangle \nonumber \\
    & =\frac{e^{\iu k\cdot r_{j}}p^{-u_{j}}z^{-v_{j}}}{\sqrt{N_{a_{\rm repr}}}}\sum_{j}h_{j}\left(a_{{\rm repr}}\right)\sum_{r=0}^{N-1}\sum_{u=0}^{1}\sum_{v=0}^{1}e^{-\iu k\cdot\left(r+r_{j}\right)}\mathcal{T}^{r+r_{j}}\left(p\mathcal{P}\right)^{u+u_{j}}p^{-u_{j}}\left(z\mathcal{Z}\right)^{v+v_{j}}z^{-v_{j}}\vert b_{{\rm repr}}^{j}\rangle \nonumber \\
    & =\sum_{j}\frac{\sqrt{N_{b^{j}_{\rm repr}}}}{\sqrt{N_{a_{\rm repr}}}}h_{j}\left(a_{{\rm repr}}\right)e^{\iu k\cdot r_{j}}p^{-u_{j}}z^{-v_{j}}\vert b_{{\rm symm}}^{j}\left(k,p,z\right)\rangle,
\end{align}
where $\vert b_{{\rm real}}^{j}\rangle=\mathcal{T}^{r_{j}}\mathcal{P}^{u_{j}}\mathcal{Z}^{v_{j}}\vert b_{{\rm repr}}^{j}\rangle$, and $N_{b^{j}_{\rm repr}}$ is the normalization factor of $\vert b_{{\rm symm}}^{j}\left(k,p,z\right)\rangle$.

Ultimately, the Hamiltonian matrix elements in the symmetric basis are given by:

\begin{equation}
\langle b_{{\rm symm}}^{j}\left(k,p,z\right)\vert \mathcal{H}\vert a_{{\rm symm}}\left(k,p,z\right)\rangle=\frac{\sqrt{N_{b^{j}_{\rm repr}}}}{\sqrt{N_{a_{\rm repr}}}}h_{j}\left(a_{{\rm repr}}\right)e^{\iu k\cdot r_{j}}p^{-u_{j}}z^{-v_{j}}.
\end{equation}

\section{Truncation of the RDM Subspace}
\label{app:weight}
\begin{figure}[h]
    \centering
    \includegraphics[width=\textwidth]{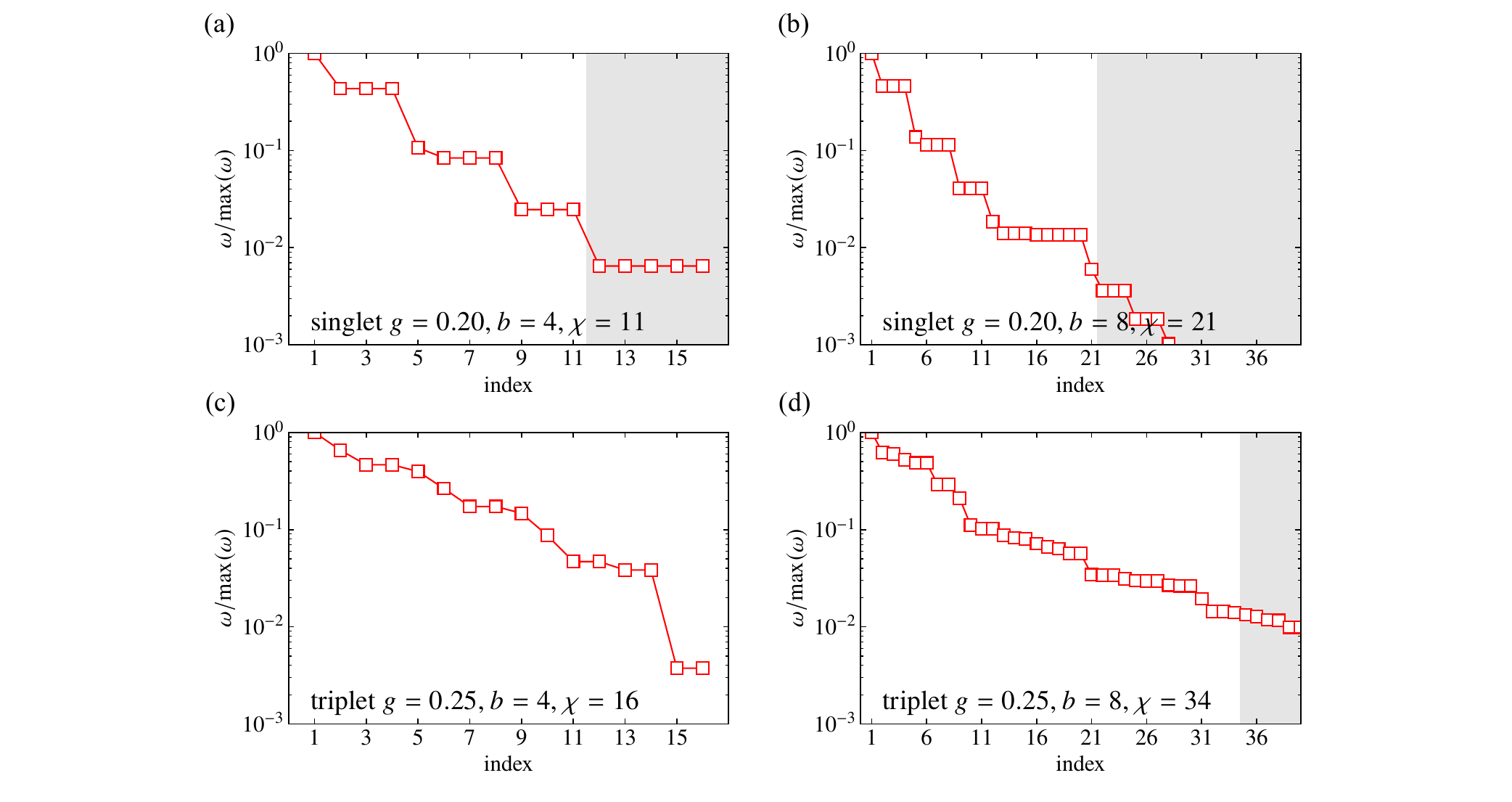}
    \caption{Eigenvalues $\omega$ of the RDM $\rho_{b}$, rescaled by the largest eigenvalue and shown in descending order, obtained from the singlet and triplet states of the $N=16$ system at $g=0.2$ and $0.25$. (a) Singlet, $g=0.2$, $b=4$, $\chi=11$. (b) Singlet, $g=0.2$, $b=8$, $\chi=21$. (c) Triplet, $g=0.25$, $b=4$, $\chi=16$. (d) Triplet, $g=0.25$, $b=8$, $\chi=34$. In panels (b) and (d), only the first 40 of the 256 normalized eigenvalues are shown. The grey regions highlight the eigenvalues discarded when the block isometry $C_i$ is truncated to $\chi$ states.}
    \label{fig:weight}
\end{figure}

The renormalized block-basis dimension $\chi$ is chosen from the eigenvalue spectrum of the reduced density matrix (RDM) $\rho_b$ of a block of $b$ spins. The RDM is obtained from a small system, here $N=16$, whose eigenstate in the target symmetry sector can be computed exactly by ED.

Figure~\ref{fig:weight} shows the eigenvalues $\omega$ of $\rho_b$, rescaled by the largest one, in descending order for the 1D $J_1\!-\!J_2$ chain in both the singlet and triplet sectors. The spectrum decays rapidly, so that nearly all of the weight of $\rho_b$ is carried by a small number of states, and a relatively small $\chi$ therefore suffices to represent the local degrees of freedom. Retaining the $\chi$ states of largest weight is precisely the truncation that defines the block isometry $C_i$ of Sec.~\ref{sec:method}; it preserves the variational accuracy while substantially reducing the number of parameters carried by the MPS backbone.

No closed-form criterion is used to fix $\chi$; the choice is empirical, and the two sectors behave quite differently. For the singlet ground state the spectrum falls off quickly and a modest $\chi$ already captures essentially all of the weight --- $\chi=11$ out of $16$ block states at $b=4$, and $\chi=21$ out of $256$ at $b=8$ --- so that the reduction becomes pronounced as $b$ grows. For the lowest triplet excitation the decay is markedly slower, and a larger $\chi$ is required at the same block size, $\chi=16$ and $\chi=34$ respectively; at $b=4$ the retained states then exhaust the block space, no eigenvalue is discarded, and $C_i$ reduces to a unitary rather than a proper isometry. We therefore scan several values of $\chi$ at fixed $b$ and $D$ and retain the one that yields the lowest variational energy, using the spectra of Fig.~\ref{fig:weight} as a guide to the range worth testing rather than as a sharp cutoff.

\section{Entanglement Entropy}
\label{app:ee}

\begin{figure}[H]
    \centering
    \includegraphics[width=0.3\textwidth]{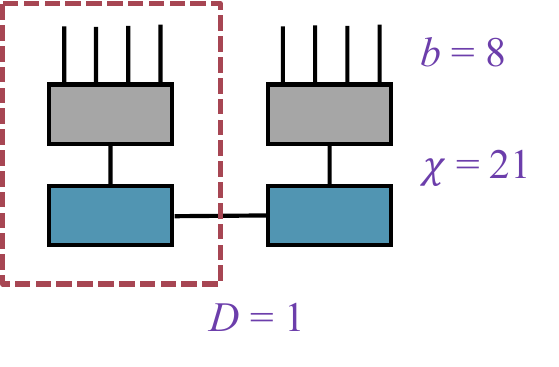}
    \caption{Tensor-network layout used for the entanglement entropy comparison. The central bond partitions the system into subsystems $A$ and $B$.}
    \label{fig:ee}
\end{figure}

To quantify how the symmetry-adapted representation reduces the entanglement burden on the MPS backbone, we compare two states that differ only in the basis in which they are defined. We consider an $N=16$ system at $g=0.2$, using the two-block layout shown in Fig.~\ref{fig:ee}, with $n_b=2$, $b=8$, $\chi=21$, and $D=1$; the EE is evaluated across the central bipartition.

Both states are built from the same generator, Eq.~\eqref{phia}. The two block isometries $C_1$ and $C_2$ are fixed to the dominant RDM eigenvectors as described in Sec.~\ref{sec:method}, so that the only variational freedom lies in the boundary tensors $B^{\gamma_1}_{\mu_1}$ and $B^{\gamma_2}_{\mu_1}$, which at $D=1$ reduce to vectors of length $\chi$; each state therefore carries the same $2\chi=42$ variational parameters. The two differ only in how the generator is used: in the symmetry-adapted case it supplies the modulus and phase of $\psi(a_{\rm symm})$ through Eqs.~\eqref{amp_psi} and~\eqref{arg_psi}, whereas in the reference case the amplitude is taken directly as the real-space wave function. The tensors $B$ are optimized independently in the two cases, each minimizing the variational energy of its own representation, so that neither state is handicapped by a choice made for the other. The reference calculation is therefore not a conventional MPS calculation and is not intended as one; it is the closest available control, sharing the ansatz, the block isometries and the parameter count, so that any difference between the two can be attributed to the basis alone. Being a real-space state, it cannot be assigned quantum numbers, whereas the symmetry-adapted state lies in the $(k,p,z)=(0,1,1)$ sector by construction. Both are compared with the exact ground state, which belongs to that sector.

For the decomposition $\mathcal{H}=\mathcal{H}_{A}\otimes\mathcal{H}_{B}$, the EE of subsystem $A$ is
\begin{equation}
S_{A} = -\operatorname{Tr} \rho_{A} \ln \rho_{A},
\label{eq:EE}
\end{equation}
where $\rho_A=\operatorname{Tr}_B|\psi\rangle\langle\psi|$ is the reduced density matrix of subsystem $A$.

Using the relation between the wave-function coefficients in the real-space and symmetric bases [Eq.~\eqref{eq:psireal}], we compute the EE from Eq.~\eqref{eq:EE}.

The results are collected in Table~\ref{ee}.
\begin{table}[H]
\caption{Entanglement entropy $S_A$ and relative energy error $\epsilon(N)$ for the exact, symmetry-adapted, and unsymmetrized real-space states.}
\label{ee}
\begin{ruledtabular}
\begin{tabular}{lcc}
State & Relative error $\epsilon(N)$ & Entanglement entropy $S_{A}$ \\
\colrule
Ground state (ED) & --- & $1.2832$ \\
First excited state in the target sector (ED) & --- & $2.0436$ \\
Symmetry-adapted state ($D=1$) & $7.54\times10^{-4}$ & $1.2895$ \\
Real-space state ($D=1$) & $2.05\times10^{-2}$ & $7.04\times10^{-9}$ \\
\end{tabular}
\end{ruledtabular}
\end{table}

At $D=1$ the single bond of the two-block layout is trivial, so the generator $\phi$ factorizes across the central partition and carries exactly zero bipartite entanglement; the same is true of the unsymmetrized real-space state, whose $S_A$ vanishes identically, the tabulated $7.04\times10^{-9}$ being finite-precision noise. That the symmetry-adapted state built on such a generator nonetheless attains $S_{A}=1.2895$, within $0.5\%$ of the exact ground-state value $1.2832$, shows that its bipartite entanglement originates entirely in the symmetric basis rather than in the tensor-network backbone. The small excess over the exact value is consistent with a residual admixture of higher-lying states of the same symmetry sector, which carry more bipartite entanglement. The corresponding relative energy error is $7.54\times10^{-4}$, against $2.05\times10^{-2}$ for the unsymmetrized state.

\end{document}